\documentclass[prd, aps, superscriptaddress, preprintnumbers, twocolumn, floatfix]{revtex4}

\usepackage{amsfonts}
\usepackage{amsmath}
\usepackage{amssymb}
\usepackage{bm}
\usepackage{dcolumn}
\usepackage{graphicx}   
\usepackage[latin1]{inputenc}
\usepackage{latexsym}
\usepackage{rotating}
\usepackage{graphicx}
\usepackage{color}
\usepackage{float}

\usepackage{amsfonts}
\usepackage{amsmath}
\usepackage{amssymb}
\usepackage{bm}
\usepackage{dcolumn}
\usepackage{epsfig}
\usepackage{graphicx}
\usepackage[latin1]{inputenc}
\usepackage{latexsym}
\usepackage{rotating}

\usepackage[unicode=true,pdfusetitle,
 bookmarks=true,bookmarksnumbered=false,bookmarksopen=false,
 breaklinks=false,pdfborder={0 0 1},backref=false,colorlinks=true]
 {hyperref}
\hypersetup{
 linkcolor=blue, citecolor=magenta, urlcolor=red, filecolor=blue}
 
\newcommand\be{\begin{equation}}
\newcommand\ba{\begin{eqnarray}}
\newcommand\ee{\end{equation}}
\newcommand\ea{\end{eqnarray}}

\newcommand{\tildet}{{\tilde{t}}}

\begin{document}

\title{T-Dual Cosmological Solutions of Double Field Theory II}

\author{Heliudson Bernardo}
\email{heliudson@gmail.com}
\affiliation{Instituto de F\'isica Te\'orica, UNESP-Universidade Estadual Paulista,
R. Dr. Bento T. Ferraz 271, S\~ao Paulo 01140-070, SP, Brazil, and\\
Department of Physics, McGill University, Montreal, QC, H3A 2T8, Canada}

\author{Robert Brandenberger}
\email{rhb@physics.mcgill.ca}
\affiliation{Department of Physics, McGill University, Montreal, QC, H3A 2T8, Canada}

\author{Guilherme Franzmann}
\email{franzmann@physics.mcgill.ca}
\affiliation{Department of Physics, McGill University, Montreal, QC, H3A 2T8, Canada}

\date{\today}

\begin{abstract}

In this paper we present cosmological solutions of Double Field Theory in the supergravity frame and in the winding frame which are related via T-duality. In particular, we show that the solutions can be viewed without the need of complexifying the cosmological scale factor.

\end{abstract}

\maketitle

\section{Introduction} 

Target space duality \cite{Tdual} is a key symmetry of superstring theory. Qualitatively speaking, it states that physics on small compact spaces of radius $R$ is equivalent to physics on  large compact spaces of radius $1/R$ (in string units). This duality is a symmetry of the mass spectrum of free strings: to each momentum mode of energy $n/R$ (where $n$ is an integer) there is a winding mode of energy $mR$, where $m$ is an integer. Hence, the spectrum is unchanged under the symmetry transformation $R \rightarrow 1/R$ if the winding and momentum quantum numbers $m$ and $n$ are interchanged. The energy of the string oscillatory modes is independent of $R$. This symmetry is obeyed by string interactions, and it is also supposed to hold at the non-perturbative level (see e.g. \cite{Pol}).

The exponential tower of string oscillatory modes leads to a maximal temperature for a gas of strings in thermal equilibrium, the Hagedorn temperature \cite{Hagedorn}. Combining these thermodynamic considerations with the T-duality symmetry lead to the proposal of {\it String Gas Comology} \cite{BV} (see also \cite{KP}), a nonsingular cosmological model in which the Universe loiters for a long time in a thermal state of strings just below the Hagedorn temperature, a state in which both momentum and winding modes are excited. This is the {\it `Hagedorn phase'}. After a phase transition in which the winding modes interact to decay into loops, the T-duality symmetry of the state is spontaneously broken, the equation of state of the matter gas changes to that of radiation, and the radiation phase of Standard Big Bang expansion can begin.

In addition to providing a nonsingular cosmology, String Gas Cosmology leads to an alternative to cosmological inflation for the origin of structure \cite{NBV}: According to this picture, thermal fluctuations of strings in the Hagedorn phase lead to the observed inhomogeneities in the distribution of matter at late times. Making use of the holographic scaling of matter correlation functions in the Hagedorn phase, one obtains a scale-invariant spectrum of cosmological perturbations with a slight red tilt, like the spectrum which simple models of inflation predict \cite{NBV}. If the string scale corresponds to that of Grand Unification, then the observed amplitude of the spectrum emerges naturally. String Gas Cosmology also predicts a roughly scale-invariant spectrum of gravitational waves, but this time with a slight blue tilt \cite{BNPV}, a prediction with which the scenario can be distinguished from simple inflationary models (see also \cite{Siyi} and \cite{Liang} for other distinctive predictions).

The phase transition at the end of the Hagedorn phase allows exactly three spatial dimensions to expand, the others being confined forever at the string scale by the winding and momentum modes about the extra dimension (see \cite{Patil, Watson} for detailed discussions of this point). The dilaton can be stabilized by the addition of a gaugino condensation mechanism \cite{Frey}, without disrupting the stabilization of the radii of the extra dimensions. Gaugino condensation also leads to supersymmetry breaking at a high scale \cite{Wei}. The reader is referred to \cite{SGCreviews} for detailed reviews of the String Gas Cosmology scenario.

However, an oustanding issue in String Gas Cosmology is to obtain a consistent description of the background space-time. Einstein gravity is clearly not applicable since it is not consistent with the basic T-duality symmetry of string theory. Dilaton gravity, as studied in {\it Pre-Big Bang Cosmology} \cite{PBB} is a promising starting point, but it also does not take into account the fact, discussed in detail in \cite{BV}, that to each spatial dimension there are two position operators, the first one ($x$) dual to momentum, the second one (${\tilde{x}}$) dual to winding. {\it Double Field Theory} (DFT) (see \cite{Siegel,HZ} for original works and \cite{Grana} for a detailed review) is a field theory model which is consistent both with the T-duality symmetry of string theory and the resulting doubling of the number of spatial coordinates (see also \cite{DFTearly} for some early works). Hence, as a stepping stone towards understanding the dynamics of String Gas Cosmology it is of interest to study cosmological solutions of DFT.

In an initial paper \cite{DFT1}, point particle motion in doubled space was studied, and it was argued that, when interpreted in terms of physical clocks, geodesics can be completed arbitrarily far into the past and future. In a next paper \cite{DFT2}, the cosmological equations of dilaton gravity were studied with a matter source which has the equation of state of a gas of closed strings. Again, it was shown that the cosmological dynamics is non-singular. The full DFT equations of motion in the case of homogeneous and isotropic cosmology were then studied in \cite{DFT3}. The consistency of DFT with the underlying string theory leads to a constraint. In DFT, in general a stronger version of this constraint is used, namely the assumption that the fields only depend on one subset of the doubled coordinates. There are various possible frames which realize this (see the discussion in the following section). In the {\it supergravity frame} it is assumed that the fields do not depend on the ``doubled'' coordinates ${\tilde{x}}$, while in the {\it winding frame} it is assumed that the fields only depend on ${\tilde{x}}$
and not on the $x$ coordinates. It was shown that for solutions with constant dilaton in the supergravity frame, the consistency of the equations demands that the equation of state of matter is that of relativistic radiation, while constant dilaton in the winding frame demands that the equation of state of matter is that of a gas of winding modes. These two solutions, however, are not T-dual. In this paper we will look for solutions which are T-dual. We expand on the analysis of \cite{DFT3} and present improvements in the solutions.

In the following section we discuss different frames which can be used. They can be obtained from each other by T-duality transformations. We also discuss the T-duality transformation of fields. In Section 3 we present the equations of DFT for a homogeneous and isotropic cosmology. In Section 4 we introduce a T-duality preserving ansatz for the solutions, before finding solutions of these equations in Section 5. We conclude with a discussion of our results.

\section{T-Dual Frames vs. T-Dual Variables}

We consider an underlying D-dimensional space-time. The fields of DFT then live in a $2D$ dimensional space with coordinates $(t, x)$ and dual coordinates $({\tilde{t}}, {\tilde{x}})$, where $t$ is time and $x$ denote the $D - 1$ spatial coordinates. In general, the {\it generalized metric} of DFT is made up of the $D-$dimensional space-time metric, the dilaton and an antisymmetric tensor field, all being functions of the $2D$ coordinates. 

In this section (like in the rest of this paper) we consider only homogeneous and isotropic space-times and transformations which preserve the symmetries. In this case, the basic fields reduce to the cosmological scale factor $a(t,  {\tilde{t}})$ and the dilaton $\phi(t, {\tilde{t}})$. It is self-consistent to neglect the antisymmetric tensor field. These are the same fields which also appear in dilaton gravity.

In supergravity, the T-duality transformation of the fields can be defined as
\ba
a(t) \, &\rightarrow& \, \frac{1}{a(t)}  \\
d(t) \, &\rightarrow& \, d(t) \, ,
\ea
where $d(t)$ is the shifted dilaton
\be
d(t) \, = \, \phi(t) - \frac{D-1}{2} \ln a(t) 
\ee
which is invariant under a T-duality transformation. In DFT this definition can be generalized to be
\ba
a(t, \tildet) \, &\rightarrow& \, \frac{1}{a(\tildet, t)}  \\
d(t, \tildet) \, &\rightarrow& \, d(\tildet, t) \, .
\ea
This implies that dilaton transforms as
\be \label{dilatontransf}
\phi(t, \tildet) \, \rightarrow \, \phi(\Tilde{t}, t) - (D-1) \ln a(\Tilde{t}, t) \, .
\ee 

An important assumption of DFT is the need to impose a {\it section condition}, a condition which states that the fields only depend on a D-dimensional subset of the space-time variables. The different choices of this section condition are called {\it frames}, and different frames are related via T-duality transformations. The {\it supergravity frame} is the frame in which the fields only depend on the $(t, x)$. The second frame which we will consider is the {\it winding frame} in which the fields only depend on the $(\tildet, {\tilde{x}})$ coordinates.

In this paper we are interested in finding supergravity frame solutions
\be
(\phi(t), a(t), d(t))
\ee
and winding frame solutions
\be
(\phi(\tildet), a(\tildet), d(\tildet))
\ee
which are T-dual to each other, i.e.
\ba
d(\tildet) \, &=& \, d(t(\tildet)) \\
a(\tildet) \, &=& \, \frac{1}{a(t(\tildet))} \, ,
\ea
where $t(\tildet) = \tildet$.

\section{Equations}

Our starting point is the equations for DFT under a cosmological ansatz \cite{DFTcosmo}  (Eqs. (8) in \cite{DFT3}):
\begin{widetext}
\begin{align}
    4d^{\prime\prime}- 4(d^{\prime})^2 -(D-1)\Tilde{H}^2+ 4\Ddot{d}-4\Dot{d}^2-(D-1)H^2 &= 0\nonumber\\
    (D-1)\Tilde{H}^2 - 2d'' -(D-1)H^2 +2\Ddot{d} &=0\nonumber\\
    \Tilde{H}' - 2\Tilde{H}d' + \Dot{H}-2H \Dot{d} &= 0 \, ,
\end{align}
where the prime denotes the derivative with respect to $\tildet$, and the overdot the derivative with respect to $t$. In addition,
\be
H \, = \, \frac{{\dot{a}}}{a} \,, \quad {\tilde{H}} \, = \, \frac{a^{\prime}}{a} \, .
\ee
These equations are invariant under T-duality, since $d(t, \Tilde{t})$ is a scalar and $H \leftrightarrow -\Tilde{H}$ under this transformation. Then, we couple these equations with matter in the following way \cite{DFT3}
\begin{align} \label{mattereq}
    4d^{\prime\prime}- 4(d^\prime)^2 -(D-1)\Tilde{H}^2+ 4\Ddot{d}-4\Dot{d}^2-(D-1)H^2 &= 0\nonumber\\
    (D-1)\Tilde{H}^2 - 2d'' -(D-1)H^2 +2\Ddot{d} &=\frac{1}{2}e^{2d}E\nonumber\\
    \Tilde{H}' - 2\Tilde{H}d' + \Dot{H}-2H \Dot{d} &= \frac{1}{2}e^{2d}P. 
\end{align}
Now, these new equations are invariant under T-duality provided $E \rightarrow - E$ and $P \rightarrow -P$. But this is exactly the case since, as explained in \cite{DFT3}, the T-dual of the energy and pressure are given by
\begin{align} \label{windingT}
    E(t, \Tilde{t}) &= -2\frac{\delta F}{\delta g_{tt}(t, \Tilde{t})} \rightarrow -2\left(-g_{tt}^2(\Tilde{t},t)\frac{\delta F}{\delta g_{tt}(\Tilde{t},t)}\right) = -E(\Tilde{t},t),\nonumber\\
    P(t, \Tilde{t}) &= -\frac{2}{D-1}\frac{\delta F}{\delta g_{ij}(t, \Tilde{t})}g_{ij}(t, \Tilde{t}) = -\frac{\delta F}{\delta \ln a(t, \Tilde{t})} \rightarrow -\frac{\delta F}{\delta \ln(1/a(\Tilde{t},t))} = -P(\Tilde{t}, t),
\end{align}
\end{widetext}
where we used $g_{tt} = 1$ for our case and assumed that the matter action in double space $F$ is $O(D,D)$ invariant. The invariance of Eqs. (\ref{mattereq})  under T-duality is a strong support for the correctness of the coupling with matter. 

Solutions to Eqs. (\ref{mattereq}) may be found after imposing the strong condition of DFT. One may impose that all functions are $\Tilde{t}$-independent or $t$-independent, corresponding to the supergravity (SuGra) or winding frames, respectively. In \cite{DFT3}, solutions based on either the SuGra or winding frames were found for the case of constant dilaton $\phi(t, \Tilde{t}) =\phi_0$. But notice that by (\ref{dilatontransf}) the dilaton transforms non-trivially under T-duality. Hence, the solutions found in \cite{DFT3} in the SuGra and
winding frames, respectively, are not T-dual to each other.  The fact that two solutions both with constant dilaton in the respective frames are not related by T-duality (or $O(D,D,)$ more generally) can be confirmed by noting that equations (12) in \cite{DFT3} obtained from (\ref{mattereq}) after assuming constant dilaton are not T-dual invariant. These equations were obtained by imposing
\ba
   2d(t, \Tilde{t}) \, &=& \,  2\phi_0 -(D-1)\ln a(t, \Tilde{t}) \\
   & \implies & \,  2\Dot{d} = -(D-1)H, \quad 2d' = -(D-1)\Tilde{H} \, , \nonumber
\ea
which is not compatible with T-duality, since $2d'$ does not transform to $2\Dot{d}$ as it should. 

From the point of view of a field theory with doubled coordinates, there is no problem in considering constant dilaton in the way it was considered in \cite{DFT3}. However, since the SuGra and winding frame solutions are not T-dual to each other, the comparison of these solutions used to motivate the correspondence $\Tilde{t} \rightarrow t^{-1}$ is tenuous.

In this work, we look for equations and solutions that respect T-duality, and specifically with constant dilaton \textit{only} in the SuGra frame or in the winding frame. We also solve an apparent inconsistency with positive energy density in the winding frame, found in \cite{DFT3}.

\section{T-duality preserving ansatz and equations for each frame}

Starting from the supergravity frame, let us look for solutions with constant dilaton. In this case
\ba
    2d(t) \, & = & \,  2\phi_0 - (D-1)\ln a(t) \nonumber \\
    & \implies & 2\Dot{d} \, = \, -(D-1)H \, .
\ea
We now seek solutions in the winding frame which are T-dual. By the invariance of $d$, $d(t) = d(\Tilde{t}(t))$, we have
\ba
    \phi_0 - \frac{D-1}{2}\ln a  \, & = &  \, \phi(\Tilde{t}) -\frac{D-1}{2}\ln a(\Tilde{t}) \\ 
    & \implies & \phi(\Tilde{t}) = \phi_0 - \frac{D-1}{2}\ln\left(\frac{a(t(\Tilde{t}))}{a(\Tilde{t})}\right) \, . \nonumber
\ea
Now by the scale-factor duality which comes from the transformation of the generalized metric, $a(t(\Tilde{t})) = 1/a(\Tilde{t})$, and so
\begin{align}
    \phi(\Tilde{t}) \, = \, \phi_0 + (D-1)\ln a(\Tilde{t}) \, ,
\end{align}
and hence
\ba
    d(\Tilde{t}) \, &=& \, \phi_0 + \frac{D-1}{2}\ln a(\Tilde{t}) \nonumber \\
    & \implies & \,  2d'(\Tilde{t}) = (D-1)\Tilde{H} \, .
\ea
Thus, the ansatz for the rescaled dilaton $d(t,\Tilde{t})$ in the winding frame will be such that
\begin{align} \label{eq10}
    2\Dot{d}(t) \, = \, - (D-1)H, \quad 2d'(\Tilde{t}) \, = \, (D-1)\Tilde{H},  
\end{align}
which is related to the supergravity frame dilaton by T-duality. Similarly, for a constant dilaton in the winding frame we have
\begin{align} \label{eq11}
    2\Dot{d}(t) \, = \, (D-1)H, \quad 2d'(\Tilde{t}) \, = \, - (D-1)\Tilde{H} \, .
\end{align}
Equations (\ref{eq10}) and (\ref{eq11}) are ansaetze compatible with T-duality between the SuGra and winding frames. 

To find the equations in each frame under these assumptions, let us consider
\begin{align} \label{ansatz}
    2\Dot{d}(t) \, = \, \alpha(D-1)H, \quad 2d'(\Tilde{t}) \, = \, \Tilde{\alpha} (D-1)\Tilde{H} \, ,
\end{align}
which takes both cases into account: for $(\alpha, \Tilde{\alpha}) = (-1,1)$ we have a constant dilaton in the SuGra frame and non-constant dilaton in the winding frame; for $(\alpha, \Tilde{\alpha}) = (1,-1)$, we have constant dilaton in the winding frame and non-constant dilaton in the SuGra frame. The case $(\alpha, \Tilde{\alpha})=(-1,-1)$ corresponds to having the dilaton constant in both frames and was considered in \cite{DFT3}. But, as already argued, this breaks the T-duality between the frames. Here, we are looking for solutions in each frame that are T-dual to each other, so we will not consider the case $(\alpha, \Tilde{\alpha}) = (1,1)$.

\begin{widetext}
Applying the section conditions, we get equations for SuGra and winding frame,
\begin{align}
    \begin{aligned}
        4\Ddot{d}-4\Dot{d}^2-(D-1)H^2 &= 0\\
        -(D-1)H^2 +2\Ddot{d} &=\frac{1}{2}e^{2d}E(t)\\
        \Dot{H}-2H \Dot{d} &= \frac{1}{2}e^{2d}P(t)
    \end{aligned}
    \qquad
    \begin{aligned}
        4d^{\prime\prime}- 4(d^\prime)^2 -(D-1)\Tilde{H}^2 &=0\\
         (D-1)\Tilde{H}^2 - 2d'' &= \frac{1}{2}e^{2d}E(\Tilde{t})\\
          \Tilde{H}' - 2\Tilde{H}d'&= \frac{1}{2}e^{2d}P(\Tilde{t})
    \end{aligned}    
\end{align}
\end{widetext}
Before solving them, notice that the energy and pressure in the winding frame are given by
\begin{align}
    \Tilde{E}(\Tilde{t}) &= -2\frac{\delta F}{\delta g_{\Tilde{t}\Tilde{t}}(\Tilde{t})} = -2\left(-g_{tt}^2(\Tilde{t})\frac{\delta F}{\delta g_{tt}(\Tilde{t})}\right) = -E(\Tilde{t}),\\
    \Tilde{P}(\Tilde{t}) &= -\frac{2}{D-1}\frac{\delta F}{\delta g_{\Tilde{i}\Tilde{j}}(\Tilde{t})}g_{\Tilde{i}\Tilde{j}}(\Tilde{t}) = -2\frac{\delta F}{\delta(a^{-2}(\Tilde{t}))}a^{-2}(\Tilde{t}) \nonumber\\
    &= \frac{\delta F}{\delta \ln a(\Tilde{t})} = -P(\Tilde{t}).
\end{align}
Thus, under T-duality, $E(t) \rightarrow \Tilde{E}(\Tilde{t})$ and $P(t) \rightarrow \Tilde{P}(\Tilde{t})$. This observation allows to reinterpret the minus sign appearing in the equation for $\Tilde{H}^2$ in \cite{DFT3}. In contrast to what happens in the SuGra frame, the energy measured in the winding frame is not simply the function $E(t, \Tilde{t})$ projected to $E(\Tilde{t})$ upon applying the section condition, but actually the negative of it. The difference appears because the definition of $E(t,\Tilde{t})$ selects the SuGra frame as a preferred frame, since $g_{\Tilde{t}\Tilde{t}}$ does not enter in this definition. As explained in \cite{DFT3}, to work only with $E(t,\Tilde{t})$ was a choice since the variations with respect to $g_{tt}$ can be written as $g_{\Tilde{t}\Tilde{t}}$ variations. But this choice selects $t$ as a preferred variable and so it is natural that the energy in the winding frame is different from $E(\Tilde{t})$. 

Using (\ref{ansatz}) in SuGra frame, we have
\begin{align}
    2\alpha \Dot{H}- H^2(\alpha^2(D-1)+1) &= 0, \nonumber \\
    \alpha \Dot{H}-H^2 &= \frac{1}{2(D-1)}e^{2d}E, \nonumber \\
    \Dot{H} -\alpha(D-1)H^2 &= \frac{1}{2}e^{2d}P,
\end{align}
which implies
\begin{align}
    H^2 &= \frac{e^{2\phi_0}a^{(\alpha+1)(D-1)}}{(D-1)(\alpha^2(D-1)-1)}\rho, \nonumber \\
    w &= -\frac{1}{\alpha}\frac{1}{D-1},\\
    \Dot{\rho} &+ (D-1)H(\rho + p ) = 0 \, . \nonumber
\end{align}
Notice that $\phi_0$ is the value of the dilaton in the frame where it is constant.

In winding frame we obtain
\begin{align}
    2\Tilde{\alpha}\Tilde{H}'- \Tilde{H}^2(\Tilde{\alpha}^2(D-1)+1) &= 0,\\
    -\Tilde{H}^2+\Tilde{\alpha}\Tilde{H}'&=\frac{1}{2(D-1)}e^{2d}\Tilde{E}, \nonumber \\
    -\Tilde{H}'+ \Tilde{\alpha}(D-1)\Tilde{H}^2 &= \frac{1}{2}e^{2d}\Tilde{P}, \nonumber
\end{align}
which are equivalent to
\begin{align}
    \Tilde{H}^2 &= \frac{e^{2\phi_0}a^{(\Tilde{\alpha}+1)(D-1)}}{(D-1)(\Tilde{\alpha}^2(D-1)-1)}\Tilde{\rho}, \nonumber \\
    w &= \frac{1}{\Tilde{\alpha}}\frac{1}{D-1},\\
    \Tilde{\rho}' &+ (D-1)\left[\frac{(D-1)-1/\Tilde{\alpha}}{(D-1)+1/\Tilde{\alpha}}\right]\Tilde{H}(\Tilde{\rho} + \Tilde{p})=0 \, , \nonumber
\end{align}
where $w$ is the equation of state parameter
\be
w \, = \, \frac{p}{\rho} \, ,
\ee
$p$ and $\rho$ being pressure and energy density, respectively.

From these equations, we conclude that the equation of state is the same in both frames regardless in which frame the dilaton is taken to be constant. For constant dilaton in the SuGra frame we obtain the equation of state of radiation, for constant dilaton in the winding frame, on the other hand, the equation of state is that of a gas of winding modes.

\section{Solutions}

Solving the equations of the previous section in the SuGra frame, we obtain
\begin{align}
    \rho(t)&\propto a^{-(D-1)+1/\alpha}(t),\\
    a(t)&\propto \left(\frac{\alpha}{2}(D-1)-\frac{1}{2\alpha}\right)^{\frac{2}{-\alpha(D-1)-1/\alpha}}t^{\frac{2}{-\alpha(D-1)-1/\alpha}},
\end{align}
while in the winding frame we get
\begin{align}
    \Tilde{\rho}(\Tilde{t})&\propto a^{-(D-1)+1/\Tilde{\alpha}}(\Tilde{t}),\\
    a(\Tilde{t})&\propto\left(\frac{-\Tilde{\alpha}}{2}(D-1)- \frac{1}{2\Tilde{\alpha}}\right)^{\frac{2}{-\Tilde{\alpha}(D-1)-1/\Tilde{\alpha}}}\Tilde{t}^{\frac{2}{-\Tilde{\alpha}(D-1)-1/\Tilde{\alpha}}}.
\end{align}
In particular, for constant dilaton in the SuGra frame, we have
\begin{align}
    \rho(t) &\propto a^{-D}(t),\qquad \Tilde{\rho}(\Tilde{t})\propto a^{-(D-2)}(\Tilde{t}),\\
    a(t)&\propto t^{2/D}, \qquad  a(\Tilde{t}) \propto \Tilde{t}^{-2/D}.
\end{align}
We see that given a radiation equation of state in both frames, the energy density in the winding frame has the same $a$ dependence as a fluid with winding equation of state. The reason for this is that in the winding frame the dilaton is not constant, and hence the relationship between equation of state and scale factor dependence of the energy density which we are used to from Einstein gravity changes.

For constant dilaton in the winding frame, we find
\begin{align}
    \rho(t) &\propto a^{-(D-2)}(t),\qquad \Tilde{\rho}(\Tilde{t})\propto a^{-D}(\Tilde{t}),\\
    a(t)&\propto t^{-2/D}, \qquad  a(\Tilde{t}) \propto \Tilde{t}^{2/D},
\end{align}
which shows that a fluid with winding equation of state has time dependence of the scale factor like radiation in the winding frame.

As we can check from the above results, we found solutions in the SuGra and winding frame which are T-dual to each other. Also, the solutions exhibit a symmetry connected with T-duality: if we change $t$ to $\Tilde{t}$ in the SuGra frame solution with constant dilaton in that frame, we get the winding frame solution with constant dilaton in the winding frame, and vice-versa.  

\section{Discussion}

In this paper we have constructed supergravity and winding frame solutions of the cosmological equations of Double Field Theory which are T-dual to each other. When the correct transformation of the energy and pressure is taken into account, there is no need for complexification of the scale factor.

Since Double Field Theory is based on the same T-duality symmetry which is key to superstring theory, one could hope that Double Field Theory could provide a consistent background for superstring cosmology, and provide a good background for String Gas Cosmology. Let us consider the background space to be toroidal. In this case, as argued in \cite{BV}, for large values of the radius $R$ of the torus (in string units), the light degrees of freedom correspond to the momenta, and the supergravity frame is hence the one in which observers made up of light degrees of freedom measure physical quantities. In contrast, for small values of $R$, it is the winding modes which are light, and hence the winding frame is the frame in which observers describe the physics. In the transition region (the Hagedorn phase) the full nature of double space will be important. It is possible that the section condition becomes dynamical \footnote{We thank Laurent Freidel for discussions on this point.}. It would be interesting in this context to explore the connection with the recent ideas in \cite{Freidel}.

\section*{Acknowledgement}
\noindent
 
The research at McGill is supported in part by funds from NSERC, from the Canada Research Chair program,
from a John Templeton Foundation grant to the University of Western Ontario and by the IRC - South Africa - Canada Research Chairs Mobility Initiative Grant No. 109684. HB would like to thank CAPES for supporting his work and McGill University for hospitality during an exchange period as a Graduate Research Trainee.

\end{document}